\documentstyle[prl,preprint,epsfig,aps]{revtex}
\draft
\tighten

\begin{document}

\title{  Fusion of light exotic nuclei
at near-barrier energies: effect of inelastic excitation}
\author{P. Banerjee, K. Krishan, S. Bhattacharya and C. Bhattacharya}
\address{Physics Group, Variable Energy Cyclotron Centre,
\\1/AF Bidhan Nagar, Calcutta - 700 064, INDIA}
\date{\today}
\maketitle

\begin{abstract}
The effect of inelastic excitation of exotic light projectiles (proton-
as well as neutron-rich) $^{17}$F and $^{11}$Be on fusion with heavy
target has been studied at near-barrier energies. The calculations have
been performed in the coupled channels approach where, in addition to
the normal coupling of the ground state of the projectile to the
continuum, inelastic excitation of the projectile to the bound excited
state and its coupling to the continuum have also been taken into consideration.
The inclusion of these additional couplings has been found to have 
significant effect on the fusion excitation function of neutron-rich $^{11}$Be
on $^{208}$Pb whereas the effect has been observed to be nominal for the
case of proton-rich $^{17}$F on the same target. The pronounced effect of the
channel coupling on the fusion process in case of $^{11}$Be is attributed to 
its well-developed halo structure. 
\end{abstract}


One of the interesting aspects of the study of nuclear reactions involving
radioactive beams is the possibility of using radioactive
projectiles to synthesize new, exotic heavy nuclei \cite{love93}. In
particular, nuclei located near the neutron or proton drip lines, for which
the valence particles are very loosely bound, give rise to interesting new 
phenomena, e.g., formation of halo structures \cite{han95}. The low binding
energies of the valence nucleons result in large sizes and thus, in 
increased probabilities for specific reaction channels such as nucleon
tranfer and fusion. The synthesis of heavier nuclei could thus be achieved
through fusion reactions induced by these exotic nuclei.

Due to the increasing availability of radioactive ion beams, many questions
concerning the effects of breakup processes on sub-barrier fusion of 
drip-line nuclei have been raised recently, both from the experimental 
\cite{taka97,rehm,sign,kola,das99,trot} and theoretical 
\cite{huss,taki,dasso94,hagi00} points of view. From studies of fusion of 
stable nuclei 
where breakup process is not so important, it is known that any coupling of
the relative motion of the colliding nuclei to nuclear intrinsic excitations
causes large enhancements of the fusion cross sections at sub-barrier 
energies over the predictions of a simple barrier penetration model. It is 
expected that the same thing happens for coupling to the breakup channel as 
well, especially for the weakly bound exotic nuclei lying close to or on the 
neutron/proton drip lines, for which the probability of dissociation prior 
to or at the point of contact is quite high. 
Therefore, for these nuclei, the cross sections 
for inclusive processes, i.e., the sum of complete and incomplete fusion 
cross sections should be enhanced when couplings to the breakup channels 
are considered. Considered from another point
of view, the presence of halo structure means a root mean square (rms) 
matter radius larger than the usual value deduced from the $r_0A^{1\over 3}$
systematics. As a consequence, the sub-barrier fusion cross section should
be enhanced since the Coulomb barrier is lowered. On the other 
hand, one could also argue intuitively that increased breakup probabilities
for these nuclei remove a significant part 
of flux and thus cross sections for complete fusion would be hindered. 

Since the halo effects have first been detected in the two-neutron halo
nucleus $^{11}$Li, most of the early investigations dealt with nuclei
involving weakly bound neutrons. This has been the trend as well insofar
as fusion reactions are concerned. Very recently fusion at near-barrier 
energies has been 
investigated theoretically for the neutron halo nuclei $^{11}$Be and $^6$He
on Pb target. Coupled channel calculations have been performed by 
discretizing in energy the particle continuum states \cite{hagi00}. The 
calculations show that the coupling to the breakup channels has two effects:
the loss of flux to the breakup channels and the dynamical modification of 
fusion potential. Their net effects differ depending on the energy region. 
At energies above the Coulomb barrier, the former effect
dominates over the latter and cross sections for complete fusion are hindered
compared to the no-coupling case. On the other hand, at sub-barrier energies
the latter effect is much larger than the former and complete fusion cross
sections are enhanced consequently. This is unlike the outcome of the
theoretical formulation by Hussein {\em et al.}, according to which the
breakup process always hinders complete fusion cross sections \cite{huss}.

Recently, a complete fusion excitation function has been measured for the
loosely bound nucleus $^9$Be on a Pb target at near-barrier energies 
\cite{das99}. The measurements show that cross sections for complete fusion
are considerably smaller at above-barrier energies compared with a 
theoretical calculation that reproduces the total fusion cross sections.
Also, the fusion cross sections for the $^{4,6}$He + $^{238}$U systems
measured by the SACLAY group seem to indicate that the breakup effects
enhance fusion cross sections at sub-barrier energies \cite{trot}. The
results of these two recent experiments are in general agreement with the 
theoretical findings in Ref.~\cite{hagi00}. 

Fusion induced by light proton-rich systems has received attention only recently
\cite{rehm}, but not to a good extent. Unlike the neutron-rich systems
the valence proton in the loosely bound proton-rich exotic nuclei has
to tunnel through the barrier resulting from the Coulomb repulsion due
to the charged core, which hinders the formation of proton halo. In fusion
reaction, this might not lead to a significant lowering of the Coulomb 
barrier when such nuclei interact with a target. Thus the observations
made in connection with fusion of neutron-rich nuclei at energies around
the Coulomb barrier may not be the same for them. The breakup probabilities of 
the exotic nuclei depend significantly on the separation
energy of the valence nucleon as well as its orbital angular momentum 
configuration \cite{raja}. Any non-zero angular momentum with respect to
the core will lead to a centrifugal barrier, which will restrict the
extent of the wave function in the coordinate space. The increase of the
separation energy of the valence nucleon further decreases the spread of
the wave function and thereby the cross sections of breakup processes in which 
the valence particle is removed from the nucleus also decrease \cite{raja}. If 
the exotic nucleus has some bound
excited state(s), a part of the flux will go to its inelastic transition(s) 
from the ground state to the excited state(s). Transition from the ground 
state into the continumm via the excited state(s) would be possible. 
All these could also affect its fusion with a target nucleus.

We find it interesting to investigate the roles of all these aspects that
could affect the fusion of an unstable, exotic nucleus with a stable, normal 
target. In the present work, we
report first calculations on fusion reactions induced by the proton drip
line exotic nucleus $^{17}$F and examine the effects of its inelastic 
excitation on the fusion cross sections. To compare and contrast the situation
with a light neutron-rich exotic nucleus, we also perform calculations for
the one-neutron halo nucleus $^{11}$Be.
We follow the same coupled channel formalism as in Ref.~\cite{hagi00}. But
unlike the calculations in this reference, we include continuum-continuum 
couplings. We assume the target to be inert in all cases, i.e., possible 
excitations of the target are neglected. Both Coulomb and nuclear effects
are included in the calculations. 
%\begin{eqnarray}
%\end{eqnarray}
In recent measurements with the proton drip line exotic nucleus $^{17}$F in 
fusion-fission reaction on Pb at energies around the Coulomb barrier no 
enhancement of the fusion cross section is observed \cite{rehm}. In Fig.~1,
we show our calculations on the fusion cross section for $^{17}$F + Pb at
near-barrier energies alongwith the data \cite{rehm}. We have done two sets 
of calculations. In the first set of calculation (left half of Fig.~1), we 
consider only the
transition from the ground state of $^{17}$F (600 keV below the breakup 
threshold) into the continuum. In the second set (right half of Fig.~1), the 
bound first excited state
situated 495 keV above the ground state is also included in the calculation.
This means that we have considered ground state (0$d_{5\over 2}$) to first 
excited state (1$s_{1\over 2}$) coupling (through quadrupole transition) and
also couplings from the first excited state to the continuum. In both the 
cases, the continuum up to 2 MeV has been considered and it has been found
that the results converge. The continuum has been discretized into 10 bins with
a bin size of 200 keV. The continuum states have been taken to be situated at
the middle of each bin. We consider transitions of multipolarity 1 and 2 
for transitions into the continuum, with $s$-, $p$-, $d$- and $f$-waves in the 
continuum, for the appropriate transitions. The nuclear part of the valence 
proton-target interaction, causing the inelastic transition/breakup of 
$^{17}$F, has been taken to be the same as 
the neutron-target interaction in Ref.~\cite{hagi00}. The nuclear part of the
ion-ion interaction potential has been taken to be equal to that of the 
neighbouring nucleus $^{16}$O on Pb at 88 MeV laboratory energy \cite{vid}. As 
far as the structure of 
$^{17}$F is concerned, we assume a single particle potential model in which the
valence proton moves in a Coulomb and Woods-Saxon potential relative to $^{16}$O
core. The depth of the Woods-saxon potential has been adjusted to reproduce 
the known one-proton separation energies for the bound states. 
The radius and diffuseness parameters of the Woods-Saxon potential have been 
taken to be 1.25 {\it fm} and 0.5 {\it fm} respectively.

There is good overall agreement of our calculations with the data. But
compared to the no-coupling case, we do not observe any noticeable change 
of the fusion cross sections when the channel coupling effects
are considered (see left and right parts of 
Fig.~1). However, with inclusion of the 1$s_{1\over 2}$ state in the 
calculation, we see a small increase in the complete fusion cross sections at
above-barrier energies (see left and right parts of Fig.~2). We have
done calculations for lighter targets also (not shown here) and found that 
the channel coupling effects are even smaller.

However, the observations are much different when fusion cross sections have
been computed for the one-neutron halo nucleus on a Pb target. In this case,
our calculations are different from those reported in Ref.~\cite{hagi00} in 
five aspects. Firstly,
we take into account the proper rms size, 2.91 {\it fm}, of the $^{11}$Be 
nucleus in its ground state. In Ref.~\cite{hagi00}, this was obtained from 
$r_0A^{1\over 3}$, with
$r_0$ = 1.1 {\it fm}. This gives a rms size of 2.45 {\it fm}, which 
underestimates the actual size by almost 20\%. Secondly, we also include the 
first excited state of $^{11}$Be, situated at 320 keV with respect to the 
ground state, in the calculation. Thirdly, unlike Ref.~\cite{hagi00}, where
the continuum up to 2 MeV has been considered, we consider the continuum up to
8 MeV, thereby ensuring convergence of the results. However, the continuum
discretization scheme is the same as that for $^{17}$F. Fourthly, 
continuum-continuum couplings are considered. Fifthly, we consider 
dipole transitions only, as these have been found
to be very much predominant in case of dissociation of $^{11}$Be \cite{naka}. 
But unlike the treatment of Ref.~\cite{hagi00}, in which only $p_{3\over 2}$ 
waves in the continuum are considered, we consider $s$-, $p$- and $d$-waves
in the continuum. The couplings, corresponding to dipole transitions, have
been taken appropriately. 

We display the results in Fig.~3 alongwith the data on total (i.e. complete 
plus incomplete) fusion cross section
on a $^{209}$Bi target \cite{sign}, the neighbouring nucleus of Pb, since the 
data on a Pb target is not available for $^{11}$Be induced fusion. When only 
couplings from the ground state
to the continuum are considered, we see significant enhancement of the 
sub-barrier complete and total fusion cross 
sections, and suppression of complete fusion cross sections at energies above 
the barrier. Above the Coulomb barrier, the total fusion cross sections
gradually become identical with the cross sections obtained in the 
no-coupling limit. With inclusion of the first excited state of $^{11}$Be in 
the calculation the total fusion cross sections remain almost the same at 
energies above the barrier. But both the complete and the total fusion
cross sections become larger below the barrier, by almost a factor of two at 
smaller energies. 
However, there is significant decrease ($\sim$21\%) of the complete fusion 
cross sections at above-barrier energies when the additional couplings with the 
first excited state are considered (see the two solid lines). Since complete 
fusion occurs when 
$^{11}$Be is absorbed by the target in its ground state as well as in the
bound excited state, this indicates that a considerable amount of flux goes
into inelastic excitation of $^{11}$Be to its first excited state. This is
in agreement with the fact that this transition is known to be one of the
strongest dipole transitions, with a $B(E1)$ value equal to 0.36$\pm$0.03
W.u. \cite{mill}.

In Fig.~3, the agreement of the total fusion cross sections
with the data is good at above-barrier energies. However, there are 
large error bars in the data, both in the horizontal and vertical directions, 
at sub-barrier energies. At low energies, there are only few data points. 
Further high-precision measurements, especially on complete fusion cross 
sections are suggested. We expect that it would reveal the importance of 
including the first excited state of $^{11}$Be in the calculation.

The difference in features of the fusion cross sections in reactions induced by
$^{17}$F and $^{11}$Be could be attributed to different structures of these two
nuclei.
The dominant ground state configuration of $^{11}$Be is $^{10}$Be$(0^+)\otimes 
\nu$(1$s_{1\over 2}$), whereas
for $^{17}$F it is $^{16}$O$(0^+)\otimes \pi$(0$d_{5\over 2}$). The valence 
proton in $^{17}$F feels the Coulomb barrier and the $\ell$=2 centrifugal 
barrier unlike the valence neutron in $^{11}$Be which interacts only via
nuclear potential
with the $^{10}$Be core and does not experience any centrifugal barrier. 
Therefore,
although the one-nucleon separation energies are almost the same (0.504 MeV
in $^{11}$Be and 0.6 MeV in $^{17}$F), breakup through one-nucleon removal
is much more favoured in case of $^{11}$Be as compared to $^{17}$F. We expect
that breakup of $^{17}$F will have little bearing on its fusion cross section
at near-barrier energies. This is also true so far as its inelastic 
transition to the first excited state is concerned. Dynamical calculations 
in Ref.~\cite{rehm} also show that the excitation/breakup probabilities of 
$^{17}$F are small enough to affect
the fusion process significantly. On the other hand, the cross sections
are very large for excitation/dissociation of $^{11}$Be on a heavy target 
\cite{naka}. Due
to the large size, there is more chance of lowering of the fusion barrier 
in case of $^{11}$Be. In fact, the $^{11}$Be rms radius is 2.91 {\it fm}, 
$\sim$20\% larger than that obtained from the systematics, and consequently 
the Coulomb barrier is about 1.6 MeV lower. However, $^{17}$F is a usual 
nucleus in this sense with an almost normal size of 3.04 {\it fm} \cite{kew}. 

In summary, we have performed coupled channel calculations of fusion cross
sections in reactions induced by the exotic nuclei $^{17}$F and $^{11}$Be on
Pb target. We include couplings to the inelastic and breakup channels of the
projectiles. For $^{17}$F induced reaction, there is no modification of the
fusion cross sections as compared with the no-coupling case. This is due to
the small breakup probabilities of $^{17}$F even on a heavy target, which
in turn is related to its structure in the ground state. Although proton-rich,
$^{17}$F is not a one-proton halo nucleus. On the other hand, $^{11}$Be is
an established one-neutron halo nucleus with a well-developed halo structure,
and consequently has large probabilities of undergoing dissociation. This
is reflected in the large enhancement of the sub-barrier complete and total
fusion cross sections, and suppression of the complete fusion cross sections
at energies above the Coulomb barrier. The suppression becomes less when 
additional couplings from the only bound excited state of $^{11}$Be are taken
into consideration. This is due to the strong dipole transition between the
ground state and the first excited state of $^{11}$Be, which also contributes
to the complete fusion cross section.

It would be interesting to check the effects of couplings to breakup channels
in fusion of a more exotic proton-rich nucleus, e.g., with a proton
halo and with breakup probabilities as large as those of $^{11}$Be. The
nucleus $^8$B, with one-proton separation energy of 0.137 MeV only,
would be an ideal candidate in this regard. These
calculations are under progress.

One of the authors (P.B.) thanks J. N. De for his encouragement during this
work. Fruitful discussions with Andrea Vitturi are gratefully acknowledged.

\newpage
\begin{figure}
\begin{center}
\mbox{\epsfig{file=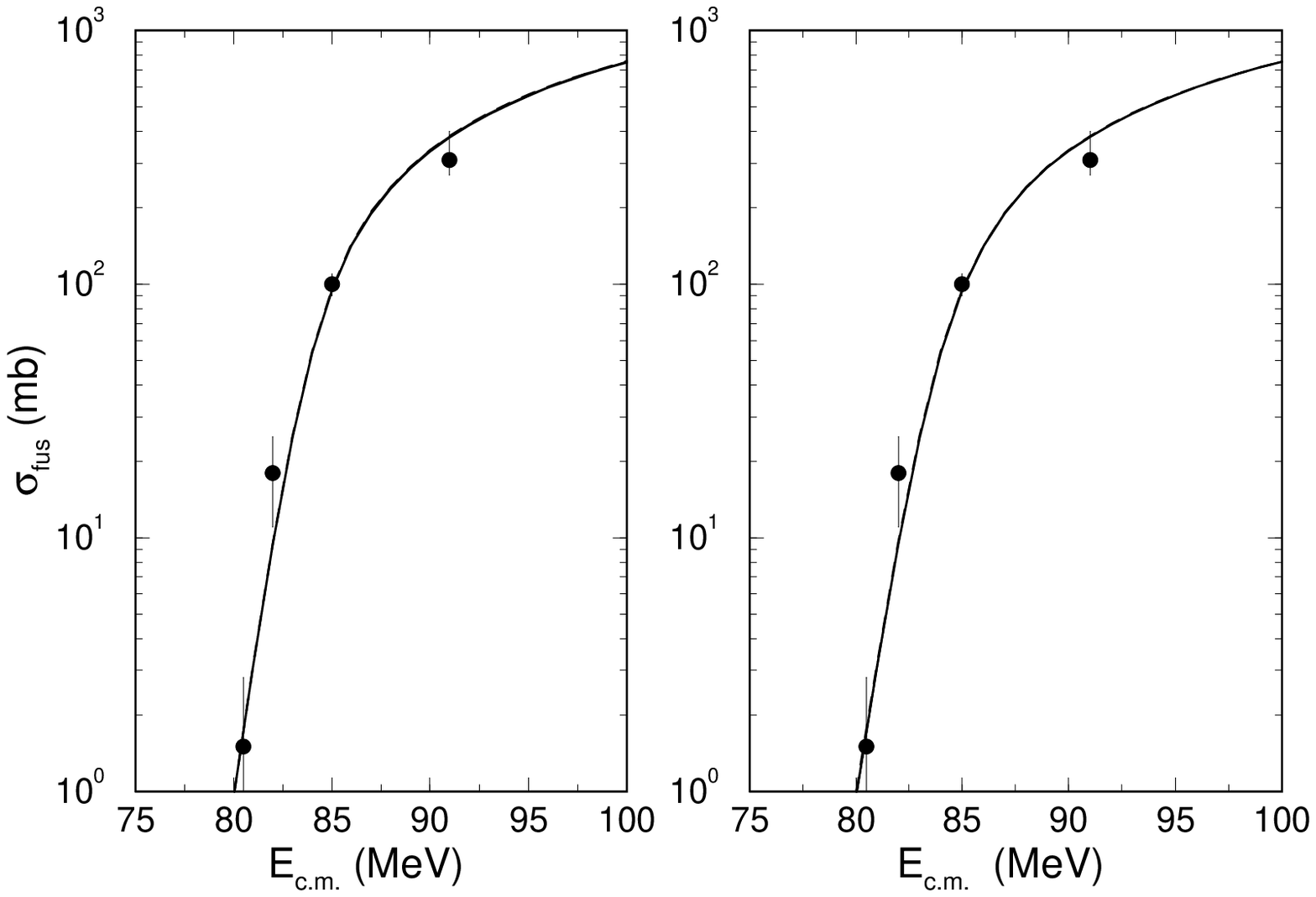,height=10cm}}
\end{center}
\caption{Fusion cross sections in the $^{17}$F + Pb reaction at energies 
around the Coulomb barrier on a semi-log scale. The thin solid line gives 
the cross sections with zero
couplings. The thick solid line and the dashed line give the complete and the
total (complete + incomplete) fusion cross sections respectively. The left part
of the 
figure shows the cross sections when only the couplings from the ground state
of $^{17}$F to the continuum are considered. The right part shows the results
when the first excited state of $^{17}$F has also been included in the 
calculation (see text). The data giving total fusion cross sections have been 
taken from \protect\cite{rehm}.}
\label{fig:figa}
\end{figure}

\begin{figure}
\begin{center}
\mbox{\epsfig{file=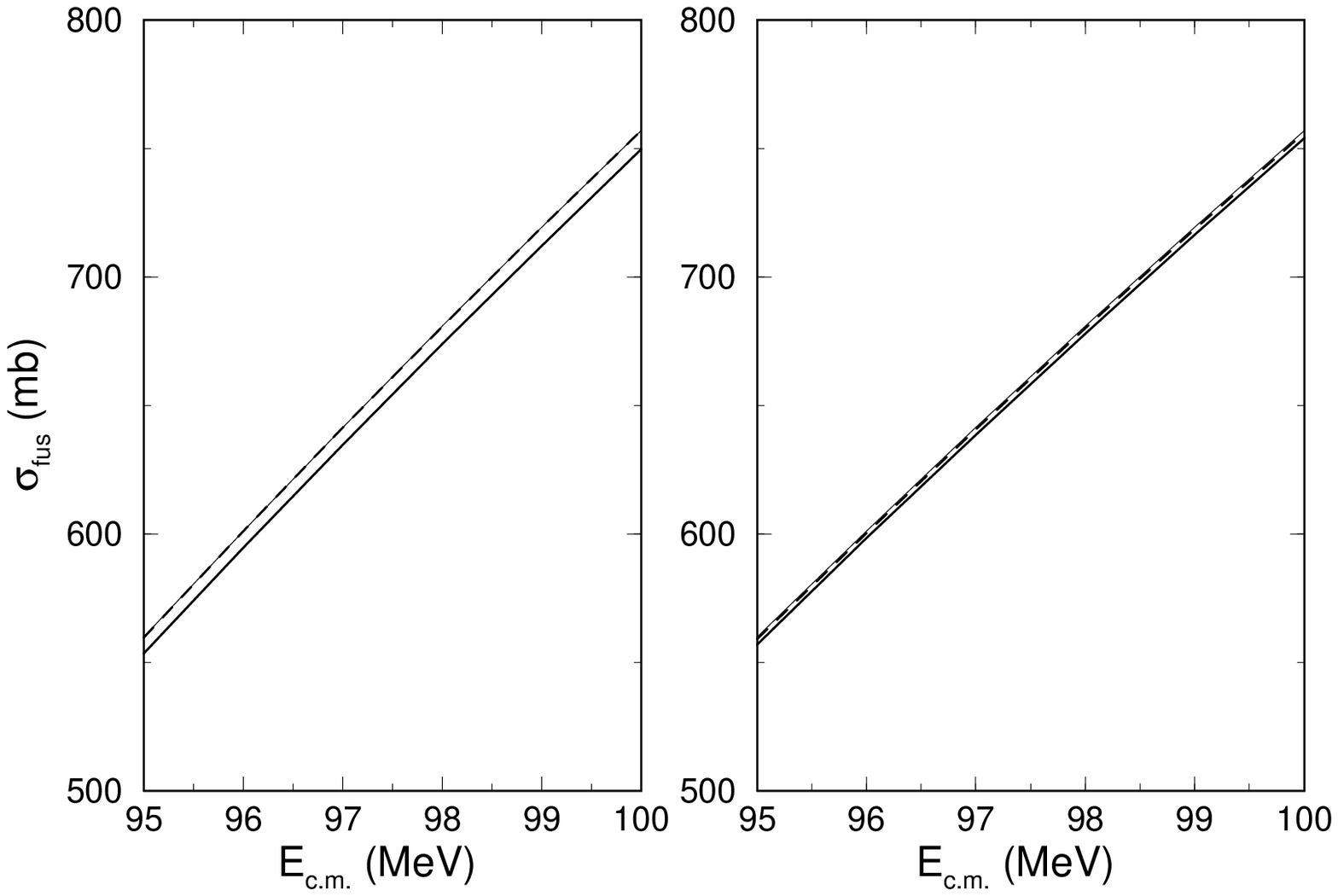,height=10cm}}
\end{center}
\caption{Fusion cross sections in the $^{17}$F + Pb reaction at energies 
above the Coulomb barrier on a linear scale. The thin solid line gives the 
cross sections with zero
couplings. The thick solid line and the dashed line give the complete and the
total (complete + incomplete) fusion cross sections respectively. The left part
of the 
figure shows the cross sections when only the couplings from the ground state
of $^{17}$F to the continuum are considered. The right part shows the results
when the first excited state of $^{17}$F has also been included in the 
calculation.} 
\label{fig:figb}
\end{figure}

\begin{figure}
\begin{center}
\mbox{\epsfig{file=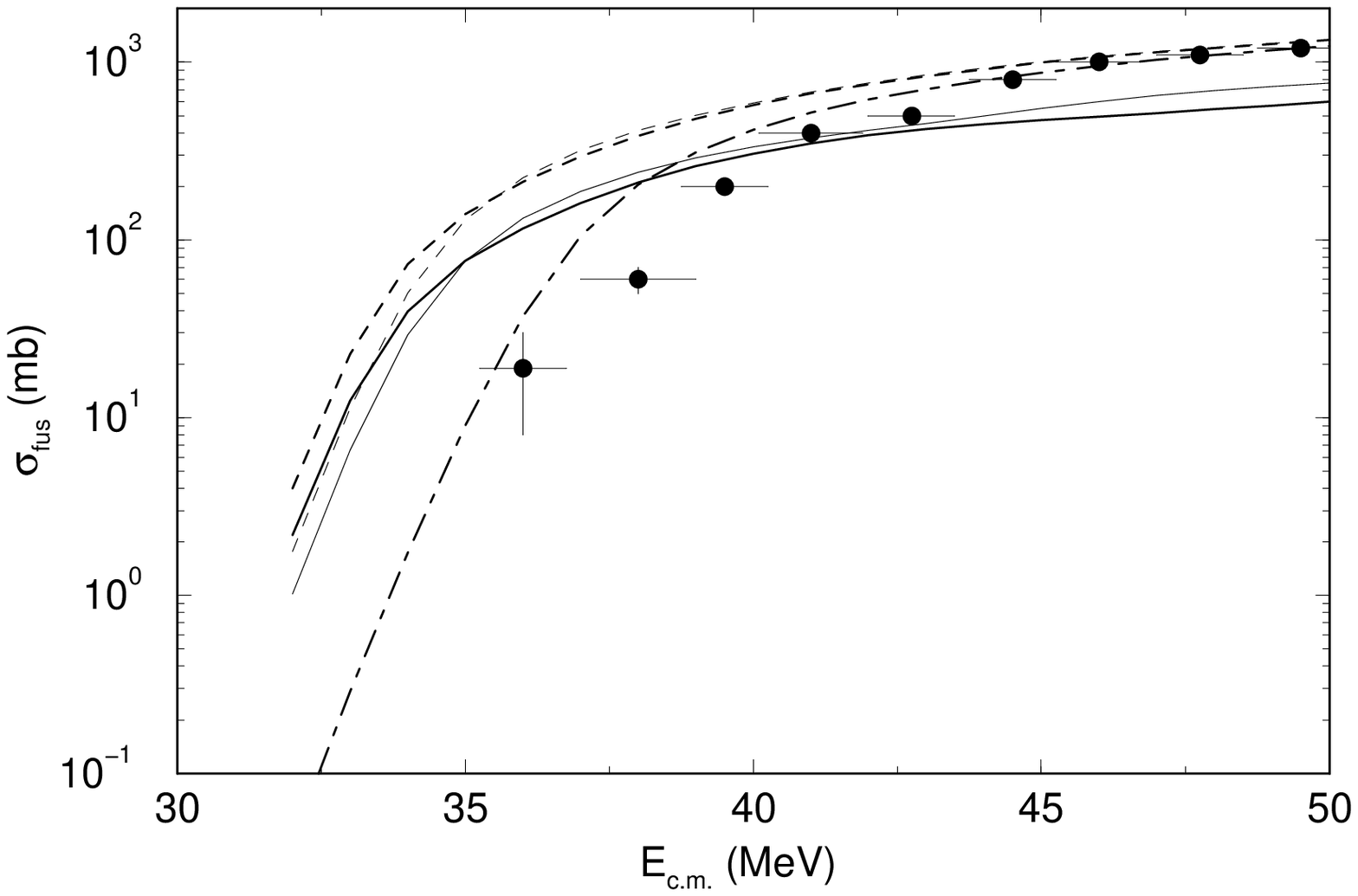,height=10cm}}
\end{center}
\caption{Fusion cross sections in the $^{11}$Be + Pb reaction at energies 
around the Coulomb barrier. The dash-dotted line gives the cross sections with 
zero couplings. The thin solid line and the thin dashed line give the complete 
and the total fusion cross sections respectively when only the couplings from 
the ground state of $^{11}$Be to the continuum are considered. 
The thick solid line and the thick dashed line show the complete and 
the total fusion cross sections respectively 
when the first excited state of $^{11}$Be has also been included in the 
calculation (see text). The data, 
taken from \protect\cite{sign}, show the total fusion cross sections on a Bi
target.}
\label{fig:figc}
\end{figure}

\begin{references}
\bibitem{love93} W. Loveland, in {\it Proceedings of the Third International
Conference on Radioactive Nuclear Beams}, edited by D. J. Morrissey (Editions
Fronti\`eres, Gif-sur-Yvette, 1993), pp. 526-536. 
\bibitem{han95} P. G. Hansen, A. S. Jensen and B. Jonson, Ann. Rev. Nucl. 
Part. Sci. {\bf 45}, 591 (1995), and references therein.
\bibitem{taka97} J. Takahashi {\em et al.}, Phys. Rev. Lett. {\bf 78}, 30 
(1997).
\bibitem{rehm} K. E. Rehm {\em et al.}, Phys. Rev. Lett. {\bf 81}, 3341 
(1998).
\bibitem{sign} C. Signiorini {\em et al.}, Eur. Phys. J. {\bf A2}, 227 (1998).
\bibitem{kola} J. J. Kolata {\em et al.}, Phys. Rev. Lett. {\bf 81}, 4580 
(1998); E. F. Aguilera {\em et al.}, Phys. Rev. Lett. {\bf 84}, 5058 (2000).
\bibitem{das99} M. Dasgupta {\em et al.}, Phys. Rev. Lett. {\bf 82}, 1395
(1999).
\bibitem{trot} M. Trotta {\em et al.}, Phys. Rev. Lett. {\bf 84}, 2342 (2000). 
\bibitem{huss} M. S. Hussein, M. P. Pato, L. F. Canto and R. Donangelo,
Phys. Rev. {\bf C46}, 377 (1992). 
\bibitem{taki} N. Takigawa, M. Kuratani and H. Sagawa, Phys. Rev. {\bf C47},
R2470 (1993).
\bibitem{dasso94} C. H. Dasso and A. Vitturi, Phys. Rev. {\bf C50}, R12 
(1994).
\bibitem{hagi00} K. Hagino, A. Vitturi, C. H. Dasso and S. M. Lenzi, Phys. 
Rev. {\bf C61}, 037602 (2000); K. Hagino and A. Vitturi, 
http://arxiv.org/nucl-th/0009013.
\bibitem{raja} R. Chatterjee, P. Banerjee, and R. Shyam, Nucl. Phys. {\bf
A675}, 477 (2000). 
\bibitem{vid} F. Videbaek {\em et al.}, Phys. Rev. {\bf C15}, 954 (1977).
\bibitem{naka} T. Nakamura {\em et al.}, Phys. Lett. {\bf B331}, 296 (1994).
\bibitem{mill} D. J. Millener {\em et al.}, Phys. Rev. {\bf C28}, 497 (1983).
\bibitem{kew} K. Krishan, P. Banerjee and R. Bhattacharya, 
http://arxiv.org/nucl-th/0109069, to be submitted to Phys. Rev. {\bf C}.
\end{references}
\end{document}